\documentclass[12pt]{article}
\title{Perturbative quantum E7(7) symmetry in N=8 supergravity}
\usepackage{mathrsfs}
\usepackage{amsmath}

\global\arraycolsep=1pt
\usepackage[colorlinks=true,backref=true,linkcolor=black,anchorcolor=black,citecolor=black,filecolor   =black,menucolor=black,pagecolor=black,urlcolor=black]{hyperref}
\usepackage[Symbol]{upgreek}
\usepackage{amsthm}
\usepackage{amssymb}
\usepackage[Symbol]{upgreek}
\usepackage{dsfont}
\usepackage{textcomp}

\usepackage{latexsym}
\usepackage{graphicx}

\usepackage{pstricks}
\usepackage{pst-node}
\usepackage{calc}
\usepackage{ifthen}

\def\bea{\begin{eqnarray}}
\def\eea{\end{eqnarray}}
\def\be{\begin{equation}}
\def\ee{\end{equation}}

\newcommand{\nn}{\nonumber}

\newcommand{\IR}{\mathbb{R}}

\newcommand{\IZ}{\mathbb{Z}}

\newcommand{\cN}{\mathcal{N}}

\newcommand{\cO}{\mathcal{O}}

\newcommand{\cI}{\mathcal{I}}

\usepackage{mathrsfs}
\usepackage[Symbol]{upgreek}
\usepackage{bbm}
\usepackage{dsfont}
\usepackage{amssymb}
\usepackage{textcomp}
\usepackage{wasysym}
\usepackage{caption}
\usepackage{graphicx}

\def\de{\delta}
\def\De{\Delta}
\def\la{\lambda}

\def\un{{\mathpzc{1}}}
\def\deux{{\mathpzc{2}}}

\newcommand{\Scal}[1]{\Bigl ({#1} \Bigr )}
\newcommand{\scal}[1]{\bigl ({#1} \bigr )}
\newcommand{\CR}{\nonumber \\*}

\def\L{{\mathcal{L}}}

\DeclareMathAlphabet{\mathpzc}{OT1}{pzc}{m}{it}

\newcommand{\ord}[1]{{\scriptscriptstyle (#1)}}

\def\ie{{\it i.e.}\ }

\def\nn{\nonumber}

\def\N{\mathcal{N}}

\def\L{{\mathcal{L}}}

\def\m{{\fontsize{10.35pt}{8pt}\selectfont {\mbox{$\mathpzc{m}$}} \fontsize{12.35pt}{12pt}\selectfont }}
\def\n{{\fontsize{10.35pt}{8pt}\selectfont {\mbox{$\mathpzc{n}$}} \fontsize{12.35pt}{12pt}\selectfont }}

\def\bm{{\fontsize{10.35pt}{8pt}\selectfont {\mbox{$\bar {\mathpzc{m}}$}} \fontsize{12.35pt}{12pt}\selectfont }}
\def\bn{{\fontsize{10.35pt}{8pt}\selectfont {\mbox{$\bar {\mathpzc{n}}$}} \fontsize{12.35pt}{12pt}\selectfont }}

\def\bigm{{\fontsize{13.35pt}{12pt}\selectfont {\mbox{$\mathpzc{m}$}} \fontsize{12.35pt}{12pt}\selectfont }}

\def\bbigm{{\fontsize{13.35pt}{12pt}\selectfont {\mbox{$\bar {\mathpzc{m}}$}} \fontsize{12.35pt}{12pt}\selectfont }}

\def\i{{\fontsize{8.35pt}{8pt}\selectfont {\mbox{\texttt{i}}} \fontsize{12.35pt}{12pt}\selectfont }}
\def\j{{\fontsize{8.35pt}{8pt}\selectfont {\mbox{\texttt{j}}} \fontsize{12.35pt}{12pt}\selectfont }}
\def\k{{\fontsize{9.35pt}{9pt}\selectfont {\mbox{\texttt{k}}} \fontsize{12.35pt}{12pt}\selectfont }}
\def\l{{\fontsize{9.35pt}{9pt}\selectfont {\mbox{\texttt{l}}} \fontsize{12.35pt}{12pt}\selectfont }}
\def\h{{\fontsize{9.35pt}{9pt}\selectfont {\mbox{\texttt{h}}} \fontsize{12.35pt}{12pt}\selectfont }}

\def\zero{{\mathpzc{0}}}
\def\un{{\mathpzc{1}}}
\def\deux{{\mathpzc{2}}}

\def\DJo{$\;$\kern-.4em \hbox{D\kern-.8em\raise.15ex\hbox{--}\kern.35em okovi\'c}}

\newcommand{\eprint}[1]{{\href{http://arxiv.org/abs/#1}{\texttt{[#1}]}}}
\newcommand{\eprintN}[1]{{\href{http://arxiv.org/abs/#1}{\texttt{#1 [hep-th]}}}}

\def\nn{\nonumber}
\def\N{\mathcal{N}}

\def\DJo{$\;$\kern-.4em \hbox{D\kern-.8em\raise.15ex\hbox{--}\kern.35em okovi\'c}}

\def\beg{\begin{equation*}}
\def\eeg{\end{equation*}}
\def\X{\mathscr{E}}
\def\ve{\varepsilon}

\allowdisplaybreaks[1]

\begin{document}

\renewcommand{\thefootnote}{\fnsymbol{footnote}}
\numberwithin{equation}{section}

\begin{titlepage}
\begin{flushright}
\
\vskip -2.5cm
{\small AEI-2011-026}\\
{\small CPHT-RR039.0511}\\
\vskip 1cm
\end{flushright}
\begin{center}
{\Large \bf Counterterms vs. Dualities}\\

\lineskip .75em
\vskip 3em
\normalsize
{\large  Guillaume Bossard\footnote{email: bossard@cpht.polytechnique.fr} and 
Hermann Nicolai\footnote{email: Hermann.Nicolai@aei.mpg.de}}\\
\vskip 1 em

$^{\ast}${\it Centre de Physique Th\'eorique, Ecole Polytechnique, CNRS\\ 91128 Palaiseau Cedex, France}\\

$^{\dagger}${\it AEI, Max-Planck-Institut f\"{u}r Gravitationsphysik\\
Am M\"{u}hlenberg 1, D-14476 Potsdam, Germany}
\\
\vskip 1 em

\vskip 3 em
\end{center}
\begin{abstract}
We investigate and clarify the mutual compatibility of the higher order corrections 
arising in supergravity and string theory effective actions and the non-linear
duality symmetries of these theories. Starting from a conventional tree level 
action leading to duality invariant equations of motion,  we show how to 
accommodate duality invariant counterterms given as functionals of both electric 
and magnetic fields in a perturbative expansion, and to deduce from them
 a non-polynomial {\em bona fide} action satisfying the Gaillard--Zumino (NGZ)
 constraint. There exists a corresponding consistency constraint in the 
non-covariant Henneaux--Teitelboim formalism which ensures that
one can always restore diffeomorphism invariance by perturbatively solving this
functional identity. We illustrate how this procedure works for the $R^2 \nabla F \nabla F$ 
and $F^4$ counterterms  in  Maxwell theory. 
\end{abstract}


\vspace{1cm}
\end{titlepage}
\renewcommand{\thefootnote}{\arabic{footnote}}
\setcounter{footnote}{0}

\pagebreak
\setcounter{page}{1}

\section{Introduction}

Consider Einstein theory coupled to scalar fields parametrising a symmetric space $G/K(G)$ 
(where $K(G)$ is the maximal compact subgroup of $G$), and $n$ abelian vector fields such that 
$G\subset Sp(2n,\mathds{R})$ acts linearly on them and their magnetic duals. This setup is typical for
the bosonic sector of various (ungauged) extended supergravity theories, and in particular for
the maximally extended $\N=8$ supergravity with duality group $G=E_{7(7)} \subset Sp(56,\mathds{R})$ \cite{CremmerJulia}. 
The purpose of this letter is to discuss the consistency of the action of this duality
group, when  higher order local corrections to the tree level action (of the type appearing 
in the string theory effective action or as counterterms in extended supergravities) are included. 
Accordingly, we will consider $n$ `electric'  vector fields $A^\m_\mu$ together with 
their `magnetic' duals $A^\bm_\mu$, combining them into a $2n$-plet of vectors 
$A_\mu^m$ with $2n$ associated field strengths $F_{\mu\nu}^m$, {\it viz.}
\be\label{Am}
A_\mu^m \equiv \big(A_\mu^\m , A_\mu^\bm\big) \quad\Leftrightarrow\quad
F_{\mu\nu}^m \equiv \big(F_{\mu\nu}^\m , F_{\mu\nu}^\bm \big)
\ee 
Note that the $n$ magnetic duals $A_\mu^\bm$ are only defined {\em on shell},
as non-local functionals of the other fields of the theory.~\footnote{In
   the literature \cite{GZ,AFZ,Kallosh} the magnetic field strengths are often denoted by the
   letter $G_{\mu\nu}$, so the relation with our notation (which follows \cite{BHN}) is established
   by making the identification $(F^\m_{\mu\nu},F^\bm_{\mu\nu}) \equiv (F_{\mu\nu}^\m,G_{\mu\nu}^\m)$.
   Because the extension of our arguments to fermions is straightforward, we will not consider
   fermions in this paper, but see {\it e.g.} \cite{Christian}.}
Classically, this redundancy is reflected in the so-called {\em twisted selfduality
constraint} \cite{CremmerJulia} for the $2n$ field strengths $F_{\mu\nu}^m$ 
\be\label{EOM2}
{F}_{\mu\nu}^m=-\frac{1}{2\sqrt{\mbox{-}g}}\ve_{\mu\nu}{}^{\sigma\rho}J^m{}_n 
{F}_{\sigma\rho}^n \; \; , 
\ee
which simultaneously halves the number of degrees of freedom and puts
the theory on-shell, in such a way that the Bianchi identities for the electric 
vectors imply the equations of motion for the magnetic vectors,
and {\it vice versa}. Here, $J^m{}_n$ is a `complex structure' built from 
the $Sp(2n,\mathds{R})$ invariant symplectic form $\Omega^{mn}$ 
and the scalar field dependent symmetric metric $G_{mn} \in G$
\be 
J^m{}_n \equiv \Omega^{mp} G_{pn} \quad\Rightarrow\quad
J^m{}_p J^p{}_n = - \delta^m_n
\ee
The indices $(\bigm,\bbigm)$ correspond to the decomposition (\ref{Am}) of the 
$2n$ vectors in a Darboux basis such that the symplectic form splits as 
\be \label{Omega}
\Omega_{\m\n}  = \Omega_{\bar \m \bar \n}= 0\; ,  \qquad 
\Omega_{\m\bar \n} = - \Omega_{\bn \m} = \delta_{\m \bar \n} \; , 
\ee 
Defining $H^{\bm\bn}$ as the inverse of $G_{\bm\bn}$ one directly
obtains from (\ref{EOM2})
\be \label{EOM3}
F^\bm_{\mu\nu} = H^{\bm\bn} \biggl( \delta_{\bn \m} \frac{1}{2\sqrt{\mbox{-}g}}\ve_{\mu\nu}{}^{\sigma\rho} F_{\sigma\rho}^\m - G_{\bn \m} F^\m_{\mu\nu} \biggr) 
\quad . \ee
The classical action is then re-obtained by solving the equation
\be\label{const}
F^\bm_{\mu\nu}
= - \delta^{\bm\m} \frac{1}{\mbox{-}g} \varepsilon_{\mu\nu\sigma\rho} \frac{\delta S}{\delta F^\m_{\sigma\rho}} \quad ,
 \ee
with the result
\be\label{S0} 
S \equiv S^\ord{0}  
= - \frac{1}{4} \int d^4 x \biggl( \sqrt{\mbox{-}g} H^{\bm\bn} \delta_{\bm \m} \delta_{\bn \n} F^{\m\, \mu\nu} F^\n_{\mu\nu} + \frac{1}{2} \varepsilon^{\mu\nu\sigma\rho} H^{\bm\bn} G_{\bn \n} \delta_{\bm\m}  F^\m_{\mu\nu}  F^\n_{\sigma\rho} \biggr) \; . 
\ee
As required, the tree level action $S^\ord{0}[F_{\mu\nu}^\m]$ depends only on
the {\em electric} vector potentials. For more general actions $S$ depending on 
the electric vector fields, the basic relations (\ref{const}) remain the same, and are usually 
referred to as {\em constitutive relations} \cite{AFZ}. As shown in \cite{GZ,AFZ} 
it is a general feature that the action itself is not duality invariant, but varies as
\be
\delta^\mathfrak{g} S [F_{\mu\nu}^\m] = 
\frac{1}{8} \int d^4 x\biggl(  \varepsilon^{\mu\nu\sigma\rho}  
X^\bm{}_\n \delta_{\bm\m} F^\m_{\mu\nu} F^\n_{\sigma\rho}  - 
\frac{4}{\mbox{-}g} \varepsilon_{\mu\nu\sigma\rho} X^\m{}_\bn \delta^{\bn\n} 
\frac{\delta S}{\delta F^\m_{\mu\nu}} \frac{\delta S}{\delta  F^\n_{\sigma\rho} }  \biggr) 
\quad , \label{GZ} 
\ee
under the duality transformations
\be\label{trans}
\delta^\mathfrak{g}  G^{mn} =  X^{m}{}_p G^{pn} + X^n{}_p G^{mp} \; , 
\qquad \delta^\mathfrak{g}  F^m_{\mu\nu} = X^m{}_n F^n_{\mu\nu} \quad .
\ee
We have omitted the superscript $^\ord{0}$ in (\ref{GZ}) because, as
shown in \cite{GZ,AFZ}, the equation (\ref{GZ}) is the consistency condition for
{\em any} action $S$ with associated duality invariant equations of 
motion.\footnote{Note that (\ref{GZ}) is required for the duality transformations 
   to make sense on the fields, and this is also valid when they are not symmetries 
   of the equations of motion, but nevertheless admit a representation on the fields 
   satisfying the equations of motion.} 
In particular, it must also hold for actions including non-linear deformations
or higher order corrections, so that the duality transformations preserve 
the constitutive relations (\ref{const}).

Suppose now that we are given a classical action $S^\ord{0}$ satisfying
these requirements, such as for instance the tree level action of $\cN=8$ supergravity, 
whose vector part is just given by (\ref{S0}) (for $G=E_{7(7)}$). In perturbation theory, this action will 
be modified by higher order counterterms and corrections whose compatibility 
with duality transformations and with (\ref{GZ}) is not immediately obvious. 
The higher order corrections to the action are only defined modulo the equations 
of motion of the classical action $S^\ord{0}$. They are generally given as functionals 
of the $2n$ electric {\em and} magnetic vectors $A_\mu^m$, that is, in the 
form $\cI^\ord{1} = \cI^\ord{1}[F_{\mu\nu}^\m , F_{\mu\nu}^\bm]$.\footnote{For clarity of notation, we will use the letter $S$
 only for `true' actions defined off shell, whereas $\cI$ denotes a general functional of both electric and magnetic fields.}
In particular, the higher order counterterms in $\cN=8$ supergravity 
appear generically in this manifestly covariant form with respect 
to the duality group in terms of vector fields transforming in the linear 
$\bf{56}$ representation of $E_{7(7)}$ \cite{HL,REK,HST}.  
When trying to express the original action together with the corrections
as an actual new action functional of the electric field 
strengths only, we are thus faced with the question what expression to
substitute for the magnetic field strengths $F^\bm_{\mu\nu}$: after all,
these will be given by non-linear and possibly non-local functionals of the 
electric vector fields (as well as the other fields) whose form is determined precisely 
by the new corrected action we are looking for. A naive guess might be to substitute the
tree level solution (\ref{EOM3}), but one quickly sees that this ansatz solves
the consistency condition (\ref{GZ}) only to first order in perturbation
theory, and fails at higher orders. In other words, it could {\it a priori} appear that
the corrected action functional gives rise to inconsistencies 
with the action of the duality transformations (\ref{trans}) \cite{Kallosh,Kallosh1}.

\section{Deformed twisted selfduality constraint}

To find the right action one must therefore adopt a different strategy,
taking a deformed version of the twisted selfduality constraint as the
starting point. Namely, given a manifestly duality covariant counterterm 
correction $\cI^\ord{1}$ depending on the $2n$ field 
strengths $F^m_{\mu\nu}$ and their derivatives, we propose to replace 
(\ref{EOM2}) by the {\em deformed twisted selfduality constraint}
\be\label{DTSDC}
F^{m}_{\mu\nu} - \frac{2}{\sqrt{\mbox{-}g}}  G^{mn}  g_{\mu\sigma} g_{\nu\rho}  
\frac{ \delta \cI^\ord{1}}{\delta F^{n}_{\sigma\rho}} =
 - \frac{1}{2\sqrt{\mbox{-}g}}\ve_{\mu\nu}{}^{\sigma\rho}  J^m{}_n \left(
F^{m}_{\mu\nu} - \frac{2}{\sqrt{\mbox{-}g}}  G^{mn}  g_{\sigma\tau} g_{\rho\omega} 
\frac{ \delta \cI^\ord{1}}{\delta F^{n}_{\tau\omega}} \right)
\ee
or, equivalently,
\be 
F^{m}_{\mu\nu}+ \frac{1}{2\sqrt{\mbox{-}g}}\ve_{\mu\nu}{}^{\sigma\rho}J^m{}_n {F}_{\sigma\rho}^n   
= \frac{2}{\sqrt{\mbox{-}g}}  G^{mn}  g_{\mu\sigma} g_{\nu\rho} 
\frac{ \delta \cI^\ord{1}}{\delta F^{n}_{\sigma\rho}} + 
\Omega^{mn} \frac{1}{\mbox{-}2g} \varepsilon_{\mu\nu\sigma\rho} \frac{ \delta \cI^\ord{1}}{\delta F^{n}_{\sigma\rho}}  \; . 
\ee  
This equation is manifestly duality invariant if $\cI^\ord{1}$ 
is a duality invariant functional. At the same time it 
achieves the required halving of the number of physical degrees of
freedom and imposes the (deformed) equations of motion. To reconstruct 
a {\it bona fide} action depending only on the physical fields (and only
the electric vector fields, in particular) and
satisfying all consistency requirements, we now have two options.
\begin{itemize}
\item We first solve (\ref{DTSDC}) for the magnetic field strengths 
          $F^\bm_{\mu\nu}$ in function of the electric field strengths $F^\m_{\mu\nu}$
          and their derivatives (as well as all other fields) as a formal power series. 
          With the resulting expression for $F_{\mu\nu}^\bm$ as a functional
          of the physical fields, we then solve (\ref{const}) in a second step
          to obtain the full corrected action functional in terms of the electric
          vectors only. This procedure manifestly preserves 
          four-dimensional space-time covariance.
\item Alternatively, we can solve (\ref{DTSDC}) for the time components 
         $(F^\m_{0\i}, F^\bm_{0\i})$, and again reconstruct the requisite action
         in a second step. The resulting action depends on the {\em spatial} electric
         {\em and} magnetic vector components $(A^\m_\i, A^\bm_\i)$, and therefore
         breaks manifest space-time covariance. Nevertheless, we will see that 
         there is a consistency condition that guarantees full space-time covariance on-shell. 
\end{itemize}
Due to the non-linear dependence of $\cI^\ord{1}$ on the magnetic field
strengths $F^\bm_{\mu\nu}$ and possibly their derivatives, the resulting corrected
action will include terms of arbitrarily high order for any kind of counterterm
correction, and this will be true in both approaches. In other words, the `initial' counterterm 
$\cI^\ord{1}$, which is usually polynomial in the field strengths and their 
derivatives, must be supplemented by an infinite string of higher order 
terms. This completion of the `initial' counterterm action will thus be 
non-polynomial, and also non-local if the initial counterterm depends on
derivatives.\footnote{Note, however,  that it will nevertheless remain local in a perturbative 
  sense, {\it i.e.} involve only a {\em finite} number of derivatives at any given order 
  in the coupling constant.} Yet, it will satisfy the consistency condition (\ref{GZ}). 

Let us illustrate these claims with a simple example from Maxwell theory,
adopting the space-time covariant approach. For this purpose we
combine the electric vector $A_\mu^\un$ with its magnetic dual
$A_\mu^\deux\equiv A_\mu^{\bar{\un}}$ into a {\em complex}
vector potential $A_\mu \equiv A_\mu^\un + iA_\mu^\deux$,
with corresponding complex field strength 
\be 
F_{\mu\nu} = F_{\mu\nu}^\un + i F_{\mu\nu}^\deux \quad , 
\ee
Electromagnetic $U(1)$ duality then acts on these fields simply
as a global phase rotation. It is furthermore easy to see that 
the original (free) Maxwell equations for $A_\mu^\un$ are recovered 
from the twisted selfduality constraint
\be
F^-_{\mu\nu} = 0\quad ,
\ee
where we define the complex selfdual and anti-selfdual
field strengths as
\be 
F^\pm_{\mu\nu} := \frac{1}{2} F_{\mu\nu} \pm\frac{i}{4\sqrt{\mbox{-}g}}  
\varepsilon_{\mu\nu}{}^{\sigma\rho} F_{\sigma\rho} \; . 
\ee 
As an example of a non-trivial deformation let us consider
the $U(1)$ duality invariant expression
\be\label{MaxwellCT}
\cI^\ord{1} = - \frac{1}{4} \int d^4 x \sqrt{\mbox{-}g}  \, g^{\rho\kappa}T^{\mu\nu\sigma\lambda} 
 \,  \nabla_\mu \bar F_{\sigma\rho} \,  \nabla_\nu F_{\lambda\kappa}\, ,  
\ee
where $T^{\mu\nu\sigma\rho}$ is the Bel--Robinson tensor 
\be 
T^{\mu\nu\sigma\rho} \equiv C^{\mu\kappa\sigma\lambda} C^\nu{}_\kappa{}^\rho{}_\lambda 
- \frac{3}{2} g^{\mu[\nu} C^{\kappa\lambda]\sigma\vartheta} C_{\kappa\lambda}{}^\rho{}_\vartheta \, , 
\ee
with the Weyl tensor $C_{\mu\nu\sigma\rho}$.  The Bel--Robinson tensor is fully 
symmetric and traceless in its four indices and is conserved modulo the vacuum 
equations of motion. With the above notation, the deformed twisted 
selfduality constraint for our Maxwell example takes the form
\be
F^-_{\mu\nu} + \nabla_\sigma T_{[\mu}{}^{\sigma\rho\lambda}  
\nabla_\rho F^+_{\nu]\lambda} = 0 \; , \label{BRTwistedSelf} 
\ee
Observe that the second term in (\ref{BRTwistedSelf}) is complex anti-selfdual 
in the indices $[\mu\nu]$, as it should be, because the Bel--Robinson tensor is 
symmetric traceless and the torsion-free covariant derivatives preserve 
complex (anti)selfduality. 

Let us now construct a manifestly diffeomorphism covariant Lagrangian 
for the deformed equations of motion in terms of the {\em real}  Maxwell field strengths 
$F^\un_{\mu\nu}$ only, following the above procedure.
To this aim we define the differential operator 
\be\label{DiffOp}
(\Delta (f))_{\mu\nu} \equiv
\Delta_{\mu\nu}{}^{\rho\sigma} f_{\rho\sigma}
:= \nabla_\kappa T_{[\mu}{}^{\kappa\lambda[\sigma}  
\nabla_\lambda \delta_{\nu]}^{\rho]} f_{\rho\sigma}\; .  
\ee 
acting on two-forms $f_{\mu\nu}$.
This operator is self-adjoint and satisfies
\be 
\frac{1}{2\sqrt{\mbox{-}g}}  \varepsilon_{\mu\nu}{}^{\kappa\lambda} \, \Delta_{\kappa\lambda}{}^{\sigma\rho} = -  \Delta_{\mu\nu}{}^{\kappa\lambda} \; \frac{1}{2\sqrt{\mbox{-}g}}  \varepsilon_{\kappa\lambda}{}^{\sigma\rho}  \;  ,
\ee
thus converting selfdual into anti-selfdual tensors, and {\it vice versa}
(this accounts for the $\pm$ sign on $F^+_{\nu\lambda}$ in the second term of (\ref{BRTwistedSelf})).
Decomposing (\ref{BRTwistedSelf}) one obtains 
\be 
\scal{ \delta_{\mu\nu}^{\sigma\rho} + \Delta_{\mu\nu}{}^{\sigma\rho} }  
F^\deux_{\sigma\rho} =  \scal{ \delta_{\mu\nu}^{\sigma\rho} - 
\Delta_{\mu\nu}{}^{\sigma\rho} } \, \frac{1}{2} \varepsilon_{\sigma\rho}{}^{\kappa\lambda} 
F^\un_{\kappa\lambda}\; ,  
\ee
and inverting the operator on the left-hand side we get
\be\label{FMaxwell}
F^\deux_{\mu\nu} =  
\frac{1}{2} \varepsilon_{\mu\nu}{}^{\kappa\lambda} 
\biggl( \delta_{\kappa\lambda}^{\sigma\rho} + 2 \sum_{n\ge1}  (\Delta^n)_{\kappa\lambda}{}^{\sigma\rho} \biggr) 
F^\un_{\sigma\rho}\; ,  
\ee
where $\Delta^n$ is the $n^{\rm th}$ power of $\Delta$.  This equation, in turn, simply follows 
as the Euler--Lagrange equation of the action
\be\label{LMaxwell} 
S = - \int d^4x \sqrt{-g} \left(
  \frac{1}{4} F^{\un\, \mu\nu} F^\un_{\mu\nu} 
 + \frac{1}{2} \sum_{n=1}^\infty  F^{\un\, \mu\nu} (\Delta^n)_{\mu\nu}{}^{\sigma\rho} 
  F^\un_{\sigma\rho} \right) \; . 
\ee
The completion of the `initial' Bel--Robinson counterterm $\cI^\ord{1}$  from
(\ref{MaxwellCT}) is thus {\em non-polynomial} and also {\em non-local}
(depending on arbitrarily high powers of the derivative operator $\nabla_\mu$).
To check the consistency condition (\ref{GZ}) we first observe that
\bea \label{dSdS}
\int d^4 x \frac{2}{\mbox{-}g} \varepsilon_{\mu\nu\sigma\rho} \frac{\delta S}{\delta F^{\un}_{\mu\nu}}  \frac{\delta S}{\delta F^{\un}_{\sigma\rho}}
 &=&   
\int d^4 x \frac{1}{2} \varepsilon^{\mu\nu\sigma\rho} \biggl( \delta_{\mu\nu}^{\kappa\lambda} + 2 \sum_{n\ge1}  \Delta^n{}_{\mu\nu}^{\kappa\lambda} \biggr) F^\un_{\kappa\lambda}  \biggl( \delta_{\sigma\rho}^{\theta\tau} + 2 \sum_{n\ge1}  \Delta^n{}_{\sigma\rho}^{\theta\tau} \biggr) F^\un_{\theta\tau} \CR
&&   \!\!\!\!\!\!\!\!\!\!\!\!\!\!\!\!\!\!\!\!\!\!\!\!\!  = \,
\int d^4 x \frac{1}{2} \varepsilon^{\mu\nu\sigma\rho} F^\un_{\mu\nu}   \biggl( \delta_{\sigma\rho}^{\kappa\lambda} + 2 \sum_{n\ge1}  (\mbox{-}\Delta)^n{}_{\sigma\rho}^{\kappa\lambda} \biggr)\biggl( \delta_{\kappa\lambda}^{\theta\tau} + 2 \sum_{n\ge1}  \Delta^n{}_{\kappa\lambda}^{\theta\tau} \biggr) F^\un_{\theta\tau} \CR
&& \!\!\!\!\!\!\!\!\!\!\!\!\!\!\!\!\!\!\!\!\!\!\!\!\!  = \,   
\int d^4 x  \frac{1}{2} \varepsilon^{\mu\nu\sigma\rho} F^\un_{\mu\nu}  F^\un_{\sigma\rho} \quad .
\eea
Because $X^\un{}_\deux = - X^\deux{}_\un$ for a $U(1)$ duality rotation,
this means that the two terms on the right-hand side of (\ref{GZ}) are the same, yielding 
twice the right-hand side of (\ref{dSdS}). Now using $\delta^{\mathfrak{u}(1)} F^\un_{\mu\nu} = X^\un{}_\deux F^\deux_{\mu\nu}$
together with the constitutive relations (\ref{const}), it is straightforward to see that (\ref{GZ}) is indeed satisfied for the completed action (\ref{LMaxwell}).

The counterterm (\ref{MaxwellCT}) is actually a simplified version of a
typical term appearing in the supersymmetric completion
of the $R^4$ counterterm arising at three loops in $\N=8$ 
supergravity \cite{Freedman}, \footnote{Note that the complete supersymmetry 
  invariant is not actually duality invariant \cite{Elvang,BHS}, however its 
  non-perturbative  completion arising in string theory is believed to be 
  $E_{7(7)}(\IZ)$ invariant (see {\it e.g.} \cite{Green}), and so it is important 
  that it transforms covariantly with respect to $E_{7(7)}$.} 
where it is proportional to (using $SL(2,\mathbb{C})$ spinor notation)
\be 
C^{\alpha\beta\gamma\delta} C^{\dot{\alpha}\dot{\beta}
\dot{\gamma}\dot{\delta}}  \nabla_{\alpha\dot{\delta}} 
F_{\beta\gamma}^{ij} \nabla_{\delta\dot{\alpha}} \bar{F}_{\dot{\beta}\dot{\gamma}\, ij} \; . 
\ee 
with the $SU(8)$ field strength $F^{ij}_{\alpha\beta} \equiv 
\sigma^{\mu\nu}_{\alpha\beta} F^{ij}_{\mu\nu}$ and its complex conjugate
$\bar{F}_{\dot\alpha\dot\beta ij}$. It is rather straightforward
to generalize the above calculation and to obtain the corresponding piece of
the corrected action of $\N = 8$ supergravity to all orders. More specifically, 
with the notation and the conventions of \cite{dWN} we get
\bea  
S &=& -  \frac{1}{2} \int d^4x \sqrt{-g}  \biggl(  
\frac{1}{2}  {\rm Re}  [2S -1]^{IJ,KL}   F^{IJ\, \mu\nu}  F^{KL}_{\mu\nu}  \\ 
&& + \;  \frac{1}{2\sqrt{-g}} \varepsilon^{\mu\nu\sigma\rho} 
{\rm Im}[S]^{IJKL} F^{IJ}_{\mu\nu} F^{KL}_{\sigma\rho} 
+ \, F^{IJ\, \mu\nu} \sum_n (\Delta^n)_{\mu\nu}{}^{\sigma\rho\, IJ,KL} 
{\rm Re} [2S -1]^{KL,PQ} F^{PQ}_{\sigma\rho}  \biggr) \nn 
\eea
where $F^{IJ}_{\mu\nu}$ are the real field strengths associated with the 28
Maxwell vectors of $\N=8$ supergravity. 
The operator $\Delta^{IJ,KL}$ is defined from (\ref{DiffOp}) by covariantizing the differential operator $\nabla_\mu$ also w.r.t. the local $SU(8)$  symmetry of $\N=8$ supergravity, such that
\bea
 \Delta _{\mu\nu}{}^{\sigma\rho\, IJ,KL} &=& {\rm Re} \Big[ K^{IJ}{}_{ij} \Delta_{\mu\nu}{}^{\sigma\rho} ( u^{ij}{}_{KL} + v^{ijKL} )  \Big]   \nn \\ 
 &&\qquad  + \,  \frac{1}{2\sqrt{-g}} \varepsilon_{\mu\nu}{}^{\kappa\lambda}   {\rm Im}\Big[  K^{IJ}{}_{ij} \Delta_{\kappa\lambda}{}^{\sigma\rho} ( u^{ij}{}_{KL} + v^{ijKL} ) \Big] 
 \eea
with
\be 
K^{IJ}{}_{ij} ( u^{ij}{}_{KL} + v^{ijKL} ) = \delta^{IJ}_{KL}  \;\; , \quad
S^{IJ,KL} = K^{IJ}{}_{ij} u^{ij}{}_{KL} 
\ee

To be sure, this argument says nothing about the deformation of
local supersymmetry that must also be taken into account when counterterms
are added to the original action of $\N=8$ supergravity. To carry out such a computation
in full and explicit detail appears beyond reach, but in the following section
we will present general arguments (based on the absence of diffeomorphism and
local supersymmetry anomalies in four dimensions) that a fully invariant deformation
simultaneously compatible with nonlinear supersymmetry and $E_{7(7)}$ exists
and can be obtained at least in principle in an order by order calculation.

\section{Non-covariant formulation with manifest duality}

In the foregoing section we showed how to restore the full
duality invariance for the corrected equations of motion. However,
being on-shell, this formalism is not directly suited for quantisation
because we cannot formulate the functional Ward identities for the
duality symmetry in that case. For that purpose one must instead make use 
of a non-covariant formulation developed by Henneaux and Teitelboim \cite{HenneauxChiral}, 
and worked out for $\cN=8$ supergravity by Hillmann \cite{Christian} (see
also \cite{BHN}). In that formalism one takes the $2n$ spatial three-vectors $A_\i^m$ as 
the fundamental fields, while their time components are only defined on-shell.
As a consequence, the action is manifestly duality invariant, but no longer
manifestly invariant under space-time diffeomorphisms. At tree level,
it takes the form 
\be \label{Action1} 
S^\ord{0}_{\scriptscriptstyle \rm vec} = \frac{1}{4}   \int d^4x \Bigl(  
 \Omega_{mn}  \varepsilon^{\i\j\k} \scal{ \partial_\zero A_\i^m  
+  N^\l F^m_{\i\l}  } F_{\j\k}^n 
-  N \sqrt{h} \, G_{mn}  
h^{\i\k}h^{\j\l} F_{\i\j}^m F_{\k\l}^n  \Bigr)   \;\; . 
\ee
It is invariant only with respect to a non-standard realisation of 
space-time diffeomorphisms (but, of course, still invariant under
spatial diffeomorphisms). 
The equation of motion for the vector fields is 
\be\label{EOM1}
\ve^{\i\j\k}\partial_\j\X_\k^m = 0 \ , 
\ee
with the abbreviation
\be\label{Xk}
\X_\i^m \equiv  \partial_\zero A_\i^m  
+  N^\j F^m_{\i\j}   -  \frac{N}{2\sqrt{h}}h_{\i\j}\ve^{\j\k\l} J^m{}_n F_{\k\l}^n     \,  . 
\ee
(recall that we neglect fermionic terms). It is invariant with 
respect to the modified diffeomorphism transformation of the vector field
\be\label{DiffA}
\delta_\xi A_\i^m \equiv  \xi^j F_{ji}^m + \xi^\zero \scal{ \partial_\zero A_\i^m   -    \X_\i^m }\; . 
\ee
Although the component $A_\zero^m$ of the vector field does not appear in the 
action, its spatial gradient can be identified from the equations of motion as 
\be \label{ClasHTem}
\partial_\i A_\zero^m = \X_\i^m \quad , 
\ee
One then expresses the Lorentz field strength $F_{\mu\nu}^m$ via $F^m_{\i\j}$ and
\be\label{F0i}
F^m_{\zero\i} = \partial_\zero A_\i^m   -   \X_\i^m \quad .
\ee
With this definition, one checks that the field strength $F^m_{\mu\nu}$ 
transforms indeed as it should with respect to diffeomorphisms 
modulo the equations of motion 
\be 
\delta F^m_{\mu\nu} = \xi^\sigma \partial_\sigma F^m_{\mu\nu} - 
2 F_{\sigma[\mu} \partial_{\nu]} \xi^\sigma +   \xi^\zero \X^m_{\mu\nu} \quad ,
\ee
where $\X^m_{\mu\nu}$ is the twisted selfdual component of the equations 
of motion, \ie 
\be 
\X^m_{\i\j} =   \Omega^{mn} \varepsilon_{\i\j\k} \frac{\delta S^\ord{0}}{\delta A_\k^n} \; \; ,
\quad \X^m_{\zero\i} = - N^\j \X^m_{\i\j} +  \frac{N}{2\sqrt{h}}h_{\i\j}\ve^{\j\k\l}
J^m{}_n \X^n_{\k\l} \quad ,
\ee
in accord with the (undeformed) twisted selfduality constraint.

Next let us consider some higher order supersymmetric invariant $\cI^\ord{1}$ defined 
on-shell as a functional of $F_{\mu\nu}^m$ and the other fields of the theory, 
which is invariant with respect to the ordinary action of diffeomorphisms. 
From this action we directly obtain the corresponding off-shell action $S^\ord{1}$
by substituting (\ref{F0i}) for the time-components $F_{\zero \i}^m$, {\it viz.}  
\be \label{SunFI} 
S^\ord{1}[F^m_{\i\j}] \equiv \cI^\ord{1}\bigl[F_{\i\j}^m \, ,\, F^m_{\zero\i} \equiv  \partial_\zero A_\i^m  -  \X_\i^m \bigr] \; .
\ee
Its variation under a time-like diffeomorphism with parameter $\xi^\zero$ 
is \footnote{The covariance under {\em spatial}
  diffeomorphisms (with parameters $\xi^\i$) is manifest.}
\be 
\delta S^\ord{1} =  \int dx^4 \Omega^{mn} \varepsilon^{\i\j\k} \xi^\zero 
\frac{\delta S^\ord{1}}{\delta F^m_{\i\j}} \frac{\delta S^\ord{0}}{\delta A^n_\k} \; . 
\ee
It follows that, at the same order, the action $S^\ord{0} + S^\ord{1}$ is 
invariant with respect to the modified variation 
\be 
\delta A^m_\i =  \xi^j F_{ji}^m + \xi^\zero \Scal{ \partial_\zero A_\i^m   
-    \X_\i^m  -  \Omega^{mn} \varepsilon^{\i\j\k} \frac{\delta S^\ord{1}}{\delta F^n_{\j\k}} }  \; . 
\ee
At this order this result is precisely the expected one: the diffeormorphism transformation 
of the vector field agrees with the ordinary transformation modulo the {\em corrected} 
equations of motion. 
Of course, in order to obtain full agreement and to establish the consistency
of the deformed action one must now complete the corrected action by
adding higher order terms, just like for the covariant deformed Maxwell
action in the previous section. That is, we must determine the full invariant
\be 
S = S^\ord{0} + S^\ord{1} + S^\ord{2} + \dots
\ee
with a corresponding all order corrected transformation of the vector fields. 
The possible obstructions in carrying out this procedure are the 
solutions of the diffeomorphism Wess--Zumino consistency conditions 
as functionals of $F_{\mu\nu}^m$ and the other fields, identified modulo 
the equations of motion \cite{HenneauxCoh}. Because the action of diffeomorphisms on 
$F_{\mu\nu}^m$ is identical to the conventional one modulo the equations 
of motion, this cohomology problem is identical to the one of identifying 
algebraic diffeomorphism anomalies in four dimensions. Consequently, the 
absence of such anomalies \cite{WittenGaume,Singer} ensures the existence of a completed action $S$ which is invariant with respect to its associated diffeomorphism action. 

Similar reasoning permits to argue that the same procedure applies to supersymmetry 
invariants. The existence of a completed action invariant with respect to supersymmetry 
relies on the absence of supersymmetry anomalies. However, the argument is less 
straightforward  in that case because the supersymmetry variation of the fermion fields 
in the Henneaux--Teitelboim formulation coincides with their supersymmetry variation 
in the conventional (covariant) formulation only modulo the `integrated' classical equations 
of motion (\ref{ClasHTem}). Although proving this is beyond the scope of this paper, we 
argue that gauge invariance implies that this subtlety does not alter the proof and 
that the cohomology groups associated to the supersymmetry anomaly are isomorphic 
in the two formalisms. Equivalently, this would imply the absence of any obstruction in
defining the all order supersymmetric action as a formal power series.

We will now show how to compute the complete action $S$ perturbatively by using the 
invariance of the action as a first order functional derivative equation.
To this aim we consider the action \footnote{The covariant derivative $\nabla_\mu F_{\i\j}^m$ 
   must be defined perturbatively. At first order it is defined from the ordinary covariant 
   derivative $\nabla_\mu$ acting on $F_{\mu\nu}^m$ as defined in (\ref{F0i}), $$
   \nabla^\ord{0}_\mu F_{\i\j}^m = \partial_\mu F^m_{\i\j}  + 2 \Gamma^\k_{\mu[\i} F^m_{\j]\k} - 2 \Gamma^\zero_{\mu[\i} ( \partial_\zero A_{\j]}^m - \X_{\j]}^m ) $$.}
\be  \label{Action2} 
S = \frac{1}{4}  \int d^4x \Bigl(  
 \Omega_{mn}  \varepsilon^{\i\j\k} \scal{ \partial_\zero A_\i^m  
+  N^\l F^m_{\i\l}  } F_{\j\k}^n \Bigr) + 
I\big[F^m_{\i\j}, \nabla_\mu F^m_{\i\j},\dots \big]  \;\;, 
\ee
where the functional $I$ depends on the vector fields via the {\em spatial} field 
strengths $F_{\i\j}^m$ and their derivatives (including time derivatives).
For any such $I$ the equations of motion of the vector fields still take the
form of a spatial divergence 
\be 
\frac{\delta S}{\delta A_\i^m} = 
- \varepsilon^{\i\j\k} \partial_\j \biggl(  \Omega_{mn}  \scal{ \partial_\zero A_\k^n  
+  N^\l F^n_{\k\l}  } - \varepsilon_{\k\l\h} \frac{\delta I}{\delta F^m_{\l\h} } \biggr) = 0 \; ,  
\ee
They are thus equivalent to the first order equation 
\be 
F_{\zero\i}^m = - N^\j F_{\i\j}^m - \Omega^{mn} \varepsilon_{\i\j\k}  \frac{\delta I}{\delta F^n_{\j\k} } \; . \label{firstorder} 
\ee
Under the diffeomorphisms 
\be\label{DiffB}
\delta_\xi A_\i^m \equiv  \xi^\j F_{\j\i}^m - \xi^\zero \biggl(  N^\j F_{\i\j}^m 
+ \Omega^{mn} \varepsilon_{\i\j\k} \frac{\delta I}{\delta F^n_{\j\k}}  \biggr) \; , 
\ee
the full action $S$ transforms as 
\bea 
\delta_\xi S &=& \int d^4 x \xi^\zero \biggl( \frac{1}{4} \Omega_{mn} 
\partial_{\l} \Bigl( N^2 h^{\l\h} \varepsilon^{\i\j\k} F_{\i\h}^m F_{\j\k}^n \Bigr) 
+\Bigl( \partial_\zero F^m_{\i\j} + 2 \partial_\i N^\l F_{\j\k}^m \Bigr) \frac{\delta I}{\delta F^m_{\i\j}}  \biggr) + \delta_\xi I  \CR
&=&  \int d^4 x \xi^\zero \partial_\l \biggl( \frac{1}{4} \Omega_{mn}  N^2 h^{\l\h} \varepsilon^{\i\j\k} F_{\i\h}^m F_{\j\k}^n  
+ \Omega^{mn} \varepsilon_{\i\j\k}  \frac{\delta I}{\delta F^m_{\i\l}} \frac{\delta I}{\delta F^n_{\j\k}} \biggr) + \bar \delta_\xi I  \; , \label{ActionVariation}  
\eea
where $\bar \delta_\xi$ is defined to act on $F^m_{\i\j}$ as an ordinary diffeomorphism 
according to (\ref{firstorder})
\be 
\bar \delta_\xi F^m_{\i\j} = \xi^\mu \partial_\mu F^m_{\mu\nu} - 
2 F_{\k[\i} \partial_{\j]} \xi^\k  + 2 \biggl( N^\k F_{[\i|\k}^m + 
\Omega^{mn} \varepsilon_{\k\l[\i}  \frac{\delta I}{\delta F^n_{\k\l} } \biggr) \partial_{\j]}   \xi^\zero \; . 
\ee
The invariance of the action therefore follows from the vanishing of
\be 
\frac{\delta^L \bar \delta_\xi I }{\delta \xi^\zero} =  
\partial_\l \biggl( \frac{1}{4} \Omega_{mn}  N^2 h^{\l\h} 
\varepsilon^{\i\j\k} F_{\h\i}^m F_{\j\k}^n  + \Omega^{mn} \varepsilon_{\i\j\k}  
\frac{\delta I}{\delta F^m_{\l\i}} \frac{\delta I}{\delta F^n_{\j\k}} \biggr) \quad . \label{GZHT} 
\ee
This relation can be viewed as the non-covariant analogue of the consistency 
condition (\ref{GZ}), but now ensuring space-time covariance of our
manifestly duality invariant action. 

The equation (\ref{GZHT}) defines a functional differential equation which 
permits to determine $I$ perturbatively. This equation simplifies drastically when 
$I$ contains no explicit derivative terms, that is, when it is the integral of a 
polynomial function (`potential') $V$ of $F_{\i\j}^m$ and the metric, which is 
invariant under spatial diffeomorphisms. In that case (\ref{GZHT}) is no longer 
a functional differential equation, but reduces to the differential equation
\be  
\frac{1}{4} \Omega_{mn}  N^2 h^{\l\h} \varepsilon^{\i\j\k} F_{\h\i}^m F_{\j\k}^n  =  
\Omega^{mn} \varepsilon_{\i\j\k}  \frac{\delta I}{\delta F^m_{\l\i}} \frac{\delta I}{\delta F^n_{\j\k}} \equiv
\Omega^{mn} N^2 h \;  \varepsilon_{\i\j\k}  \frac{\partial V}{\partial F^m_{\l\i}} \frac{\partial V}{\partial F^n_{\j\k}} 
\;  .
\ee

\section{Maxwell theory, once again}

To illustrate how the procedure works in the non-covariant formulation
we again study an example generalising Maxwell theory. To keep things
as simple as possible we consider a modification that initially depends on 
the complex spatial field strength $F_{\i\j}$ polynomially, but not on its
derivatives (the inclusion of derivatives presents no problem of principle,
but renders the calculations substantially more tedious).
The tree level Lagrangian is now a function of the complex spatial
vector field $A_\i = A_\i^\un + i A_\i^\deux$ and reads
\be \L = 
- \frac{i}{2} \varepsilon^{\i\j\k} \scal{ \partial_\zero A_\i + N^\l F_{\i\l} } \bar F_{\j\k} 
+  \frac{i}{2} \varepsilon^{\i\j\k} \scal{ \partial_\zero \bar A_\i + 
N^\l \bar F_{\i\l} }  F_{\j\k} - N \sqrt{h} V[F] \;  .
\ee
with the tree level `potential'
\be 
V \equiv V^\ord{0} = h^{\i\k} h^{\j\l} F_{\i\j} \bar F_{\k\l} = F^{ab} \bar{F}_{ab} \quad .
\ee
Here and in the remainder, we will mostly use flat indices
\be
F_{ab} \equiv e_a^\i e_b^\j F_{\i\j} \,   ,
\ee
where $e_a^\i$ is the inverse dreibein such that 
$h^{\i\j}= \delta^{ab} e_a^\i e_b^\j $.
Generalising beyond tree level, the potential $V$ will be a more
 complicated function, but for any given $V$, the three vector 
transforms as
\be 
\delta A_\i = ( \xi^\j + \xi^\zero N^\j ) F_{\j\i} - i \xi^\zero 
\frac{N \sqrt{h}}{2} \varepsilon_{\i\j\k} \frac{\partial V}{\partial \bar F_{\j\k}} \; , 
\ee
In order to ensure full diffeomorphism invariance, $V$ must satisfy 
the consistency condition (\ref{GZHT}) which now reads 
\be 
\frac{\partial V}{\partial F^{a[b}} \frac{\partial V}{\partial \bar F^{cd]}} 
=\bar  F_{a[b} F_{cd]} \label{Vcons} \; . 
\ee

The general procedure then starts from some `initial' corrected potential 
of the form $V= V^\ord{0} + V^\ord{1}$ and exploits (\ref{Vcons}) in order 
to complete the potential $V$ to a more general $SO(3)$ invariant function 
of the spatial field strengths $F_{ab}$ and $\bar{F}_{ab}$, such that  
\be
V = V^\ord{0} + V^\ord{1} + \dots 
\ee
satisfies the differential equation (\ref{GZHT}). As before we 
will thus find that, for consistency, any `initial' counterterm 
$V^\ord{1}$ must be supplemented by an infinite string of higher order 
corrections. As the simpest possible example we will consider the
manifestly duality invariant expression $V^\ord{1} \propto 
\frac{1}{2} F_{\alpha\beta} F^{\alpha\beta}F_{\dot\alpha\dot\beta} F^{\dot\alpha\dot\beta}$
obtained by squaring the complex selfdual and anti-selfdual field strengths. In 
the present approach this invariant can be identified with one half the duality invariant
\be 
Y \equiv  F^{ab} F_{ab} \bar F^{cd} \bar F_{cd} 
\ee
by using the equations of motion. Writing also
\be
X \equiv F^{ab} \bar{F}_{ab}\; , 
\ee
we would thus like to solve (\ref{Vcons}) for 
\be 
V(X,Y) = F^{ab} \bar F_{ab} + 
\frac{ 1 }{2} F^{ab} F_{ab} \bar F^{cd} \bar F_{cd} + \mathcal{O}(F^6) 
\equiv X +   \frac{1}{2} Y  + \mathcal{O}(F^6)  \quad .
\ee
First of all  we note that (\ref{Vcons}) is trivially satisfied at first order because 
\be 
F_{a[b} F_{cd]} = 0 \; . 
\ee
After some further computation it is seen that $V$ must be of the form 
\be  
V(X,Y)  = X + \sum_{n=0}^{\infty}  \frac{1}{(2+2n)!} 
H^\ord{n}(X)  Y^{1+n}  \quad . \ee
The condition (\ref{Vcons}) is satisfied provided 
(using $\bar{F}_{a[b}F_{cd]} = - F_{a[b} \bar{F}_{cd]}$)
\be 
\left ( 1 +  \sum_{n=0}^{\infty}  \frac{  Y^{1+n} }{(2+2n)!} 
\frac{\partial H^\ord{n}}{\partial X\; } \right)^2 =  
Y \left(  \sum_{n=0}^{\infty}  \frac{ Y^{n} }{(1+2n)!} H^\ord{n} \right)^2 + 1  \quad .
\ee
Observe that this ansatz is manifestly duality 
invariant. At first order in $Y$ one gets 
\be 
 \frac{\partial H^\ord{0}}{\partial X\; }  = \scal{ H^\ord{0} }^2 \quad , \ee
which implies (with the condition $H^\ord{0}(0) =1$) that 
\be 
H^\ord{0}( X) = \frac{1}{1-   X} \quad .
\ee 
At order $Y^2$ we get
\be 
\frac{\partial H^\ord{1}}{\partial X\; }  - \frac{4}{1-  X} H^\ord{1} 
= - \frac{3}{(1-   X)^2} \quad , 
\ee
which gives 
\be 
H^\ord{1}(X) = \frac{1}{1- X} + \frac{c^\ord{1}}{(1- X)^4} \quad , 
\ee
with an arbitrary constant $c^\ord{1}$. This constant corresponds 
to the freedom of adding the on-shell invariant
\be 
\frac{c^\ord{1}}{24}  \scal{ F_{ab} F_{ab} \bar F_{cd} \bar F_{cd} }^2 
\approx  \frac{c^\ord{1}}{24} \scal{ F_{\alpha\beta} F^{\alpha\beta}
F_{\dot\alpha\dot\beta} F^{\dot\alpha\dot\beta} }^2 \quad , 
\ee
to the invariant $F^2\bar F^2$, while preserving diffeomorphism invariance. 

It is now clear how to proceed perturbatively in $Y$ and how to determine 
all the functions $H^\ord{n}$ by successively solving the hierarchy 
of first order equations \footnote{Where $C_n^p \equiv \frac{n!}{p!(n-p)!}$ are the binomial coefficients.} 
\bea \hspace{-1mm}
\frac{\partial \; }{\partial X } \Scal{ (1- X)^{2+2n} H^\ord{n}  }   &=&   \nonumber\\ 
&&\!\!\!\!\!\!\!\!\!\!\!\!\!\!\!\!\!\!\!\!\!\!\!\!\!\!\!\!\!\!\!\!\!\!\!\!\!\!\!\!\!\!\!\!\!\!\!\!\!\!\!\!\!\!\!
\frac{1}{2}  (1- X)^{2+2n} \left(  \sum_{p=1}^{n-1} C_{2+2n}^{1+2p} H^\ord{p} H^\ord{n-p} 
-  \sum_{p=0}^{n-1} C_{2+2n}^{2+2p} \frac{\partial H^\ord{p}}{\partial X\; }
\frac{\partial H^\ord{n-p-1}}{\partial X\; } \right) \quad .
\eea
By construction the right-hand side is a finite Laurent series in 
$(1- X)$ with polynomial coefficients in $\ln(1- X)$ which can 
be integrated straightforwardly, modulo the definition of the homogenous solution 
\be 
H^\ord{n} = \frac{c^\ord{n}}{(1-  X)^{2+2n}} +  \tilde H^\ord{n}  \quad , 
\ee
$\tilde H^\ord{n}$ being a particular solution. Clearly, the constants 
$c^\ord{n}$ correspond to the ambiguities in defining a diffeormorphism 
invariant associated to the possibility of adding higher order invariants 
corresponding on-shell to 
\be 
\frac{c^\ord{n}}{ (2n)!} \scal{ F_{\alpha\beta} F^{\alpha\beta}}^n 
\scal{ F_{\dot\alpha\dot\beta} F^{\dot\alpha\dot\beta} }^n  
\approx \frac{c^\ord{n}}{ (2n)!} \scal{ F_{ab} F_{ab}}^n \scal{ \bar F_{cd} \bar F_{cd} }^n 
\; . 
\ee
Setting $c^\ord{1} = 5 $, the potential $V$ reads 
\be 
V(X,Y) = X + \frac{1}{2} Y \scal{ 1 +  X  + X^2 }  +  \frac{1}{4} Y^2   +  
\mathcal{O}(F^{10})  \; . \label{Vorder4}  
\ee

To establish the link of the above construction with the deformed
twisted self-duality constraint (\ref{DTSDC}), we note that, by
construction, the equations of motion are invariant with respect to 
diffeomorphism invariance. Hence they can indeed be rewritten in 
this manifestly diffeomorphism covariant form (for $c^\ord{1}=5$ and higher $c^\ord{n}$ chosen appropriately), {\it viz.}
\be 
F_{\mu\nu} - \frac{i}{2\sqrt{\mbox{-}g}} \varepsilon_{\mu\nu}{}^{\sigma\rho} F_{\sigma\rho} + \frac{1}{8} \Scal{ F_{\kappa\lambda} F^{\kappa\lambda}  + \frac{i}{2\sqrt{\mbox{-}g}} \varepsilon^{\kappa\lambda\theta\tau} F_{\kappa\lambda} F_{\theta\tau} }  \Scal{ \bar F_{\mu\nu} - \frac{i}{2\sqrt{\mbox{-}g}} \varepsilon_{\mu\nu}{}^{\sigma\rho} \bar F_{\sigma\rho} } = 0 \; .  
\ee
Indeed, decomposing the corresponding equations into space and time components 
\be 
F_{\zero a} - \frac{i}{2} \varepsilon_{abc} F_{bc} +  \frac{1}{8} \scal{ F_{de} F_{de} - 2 F_{\zero d} F_{\zero d} - 2i \varepsilon_{def} F_{\zero d} F_{ef}  }  \Scal{ \bar F_{\zero a} - \frac{i}{2} \varepsilon_{abc} \bar F_{bc}} = 0 \; , 
\ee
one can perturbatively solve for $F_{\zero a}$ in terms of $F_{ab}$ as 
\begin{multline}  F_{\zero a } = \frac{i}{2} \varepsilon_{abc} F_{bc}  \left( 1 + \frac{1}{2}  Y  \scal{  1 + 2   X   } \right) \\* + \frac{i }{2} \varepsilon_a{}^{bc} \bar F_{bc} F_{ef} F_{ef} \Scal{ 1 +  X  +   X^2  +    Y } + \mathcal{O}(F^{10}) \; , \end{multline} 
This solution coincides with the expression following from the 
corrected potential $V$ obtained above in  (\ref{Vorder4}) up to order $F^{10}$. 

\section{Conclusions}

We have demonstrated for some typical examples by rather explicit computations 
that the higher order counterterms and corrections arising in supergravity 
and the effective string theory action are perfectly  compatible with the 
full non-linear duality symmetries of these theories, provided one completes the 
`initial' correction terms by solving the requisite consistency conditions. 
This can be done in either of two different formulations, in one of which
space-time covariance is manifest but the duality symmetry is realised only 
on-shell, while it is the converse in the second formulation. We have exhibited
the analogue of the Gaillard--Zumino constraint for the Henneaux--Teitelboim
formulation, and we have furthermore shown that the two procedures
give results which agree at lowest non-trivial orders in a perturbative 
expansion.

We have shown that the absence of diffeormorphism anomalies in four dimensions 
implies that one can always construct a diffeomorphism invariant corrected action functional associated to an `on-shell' invariant in the Henneaux--Teitelboim formulation. We argued 
that the absence of supersymmetry anomalies similarly implies that one can always construct a supersymmetric corrected action functional associated to any given  `on-shell'  invariant. 

We conclude that the non-linear $E_{7(7)}$ symmetry is not sufficient to rule
out all higher order counterterms, hence divergences,  of $\cN=8$ supergravity. 
At this stage of our understanding of the theory, there is unfortunately no `royal path'  to finiteness cutting short explicit calculations 
of the type performed in \cite{Bern}.  If $\cN=8$ supergravity is UV finite to all orders 
the reason must be sought beyond maximal supersymmetry and $E_{7(7)}$.

\vspace{0.5cm}

\noindent{\bf Acknowledgments:} H.~Nicolai would like to thank the Simons Center at 
Stony Brook and its staff for hospitality and support during part of this work. G.~Bossard 
is grateful to AEI Potsdam for hospitality. We would furthermore like to thank R.~Kallosh 
for discussions and comments motivating this work.


\section{Appendix: Deforming the NGZ identity}

In this appendix we discuss the NGZ identities required for the invariance of 
the equations of motion under duality \cite{GZ,AFZ,Kallosh} and their deformation
when the original action is deformed by higher order corrections. For ease of comparison 
we adopt the notation and conventions of \cite{Kallosh} throughout this appendix. 
Suppressing the dependence on other fields as well as all indices for simplicity, we write 
the action as $S=S[F]$, where, as everywhere else in this paper, the term `action' always 
refers to a functional depending on $n$ electric vectors and their associated field strengths 
$F$  only. We then define the magnetic field strengths via
\be\label{G}
\tilde{G}[F] := 2 \, \frac{\de S[F]}{\de F}
\ee
as functionals of $F$ in the standard way. If the 
equations of motion are to be duality covariant under
\be\label{ABCD}
\De F = AF + BG \quad, \quad \De G = CF +DG
\ee
with the usual $Sp(2n,\IR)$ matrix relations $A^T=-D, B=B^T$ and $C=C^T$,
the action must satisfy the NGZ constraint \cite{GZ}
($\equiv$ eq.~(3.6) in \cite{Kallosh}))
\be\label{NGZ1}
\frac{\de}{\de F} \left( S[F'] - S[F] -
\frac14 \int \big( \tilde{F}CF + \tilde{G}BG\big)\right) = 0
\ee
This is necessary for the compatibility of the transformations
(\ref{ABCD}) with the constitutive relations (\ref{G}), which  imply
\be\label{comp}
\De G[F] = \int \frac{\de G}{\de F} \De F
\ee
Now assume that these conditions are satisfied for some initial (usually
the tree level) action $S_0[F]$ with corresponding
\be\label{G0}
\tilde{G}_0[F] := 2 \, \frac{\de S_0[F]}{\de F}
\ee
We wish to `deform' this action and investigate
how the duality symmetry is deformed with it. To this aim
we expand the full action $S$ as 
\be\label{Expansion}
S[F;\la] = S_0[F] + \la S_1[F] + \cO(\la^2)
\ee
with some expansion parameter $\la$. 
Similarly,
\be
\tilde{G}[F;\la] := 2\, \frac{\de S[F;\la]}{\de F} 
= \tilde{G}_0[F] + \la \tilde{G}_1[F] + \cO(\la^2)
\ee
such that
\be\label{Gn}
\tilde{G}_k[F] := 2 \, \frac{\de S_k[F]}{\de F}
\ee
for the $k$-th order correction to the magnetic field 
strengths. The expansion of the variation is
\be
\De(\la) = \De_0 + \la\De_1 + \cO(\la^2)
\ee
Inserting this into (\ref{ABCD}) we get
\be
(\De_0 + \la\De_1 +...)F = AF + B(G_0 + \la G_1 +...)
\ee
or
\be\label{Den}
\De_0 F = AF + BG_0 \;\; , \quad \De_k F = B G_k
\ee
and
\be
(\De_0 + \la\De_1 + ...)(G_0 + \la G_1 + ...)  =\,   CF + D(G_0 + \la G_1 + ...)
 \ee
Similarly we can expand the NGZ relation (\ref{NGZ1}) 
\bea\label{NGZ2}
\frac{\de}{\de F}  &\bigg\{& \! S_0\big[F + \De F\big] 
    + \; \la S_1\big[ F + \De F\big] 
  - S_0[F] - \la S_1[F] - \dots -\frac14\int \tilde{F}CF    \nn \\
      &&  \qquad
- \frac14 \int \big( \tilde{G}_0 + \la \tilde{G}_1 + ...\big) B
       \big( G_0 + \la G_1 + ...\big)\bigg\} = 0    \nn
\eea
By assumption this relation holds at 0-th order in $\la$.
At first order in $\la$ we get 
\be
\int \left( \frac{\de S_0}{\de F} \De_1 F + \frac{\de S_1}{\de F} \De_0 F
    - \frac12 \tilde{G}_0 BG_1 \right) 
    = \int \frac{\de S_1[F]}{\de F} \De_0 F \equiv  \De_0 S_1 
\ee
where we used (\ref{G0}) and (\ref{Den}). Therefore, to this order
duality can be maintained if and only if the first order correction to the action 
is invariant under undeformed duality, or
\be
\frac{\de}{\de F} \Big( \De_0 S_1[F] \Big)= 0 \;\; \Rightarrow \;\; 
\De_0 S_1[F]= 0.
\ee
This can be arranged
by starting from a {\em manifestly duality invariant} functional 
$\cI = \cI[F,G]$ obeying
\be\label{dI}
\De\cI =\int\left(\frac{\de\cI}{\de F}(AF + BG) + \frac{\de\cI}{\de G}(CF + DG) \right)= 0
\ee
for arbitrary (hence independent) $F$ and $G$, but in particular
also for the special choice $G=G_0[F]$ and $\De = \De_0$. 
If we now take
\be
S_1[F]  := \cI\big[ F, G_0[F]\big]
\ee
the 0-th order variation comes from the varying $F$ alone 
\be
\De_0 S_1[F] = \int\left( \frac{\de\cI\big[F,G_0(F)\big]}{\de F} +
      \frac{\de\cI\big[F,G_0(F)\big]}{\de G_0}
          \frac{\de G_0}{\de F}\right) \De_0 F   
\ee
Using (\ref{comp}) together with (\ref{dI}) we see that indeed $\De_0 S_1 =0$ 
as required, and duality is maintained at $\cO(\la)$. Therefore, the first order
correction is always invariant under the {\em undeformed} duality, as borne
out also by the examples from $\N=8$ supergravity.

Up to this point our analysis agrees with \cite{Kallosh} . At second order, however, we 
disagree with the conclusions of \cite{Kallosh}. Collecting all terms of order 
$\cO(\la^2)$ in (\ref{NGZ2}) we obtain
\bea\label{Delta2}
&&\frac{\de}{\de F} \bigg\{ 
\int \bigg( \frac{\de S_0}{\de F} \De_2 F + \frac{\de S_1}{\de F} \De_1 F 
 + \frac{\de S_2}{\de F} \De_0 F  - \; \frac14 \tilde{G}_1 BG_1 - \frac12 \tilde{G}_2 BG_0 
\bigg) \bigg\} \nn\\
&& \qquad\qquad  = \frac{\de}{\de F} \left\{
\int \left( \frac{\de S_2}{\de F} \De_0 F +\frac14 \tilde{G}_1 BG_1  
\right)\right\} = 0 
\eea
(we drop all terms of higher order in any of the $\De$'s because 
the matrices $A,B,C,D$ are assumed to be infinitesimal).
In this derivation we made use again of (\ref{Gn}) and (\ref{Den})
in order to have only one unknown quantity ($= S_2$) in this expression. It is 
now clear that we must include a second order term $S_2\neq 0$
(and correspondingly higher order corrections) to salvage duality invariance
at this order (and higher orders). Furthermore, unlike the first order correction,
$S_2$ can{\em not} be invariant under undeformed duality,
but must break undeformed duality in such
a way as to cancel the second term on the r.h.s. of (\ref{Delta2}); that is,
we {\em must} have $\De_0 S_2 \neq 0$. This also implies
\be
\De_0\big(\la S_1 + \la^2 S_2 + \dots) = \cO(\la^2) \neq 0
\ee
so (3.9) of \cite{Kallosh} (stating $\De_0 \hat{S} =0$) is only correct to 
first order in $\la$, but fails beyond.  Let us finally note that, assuming
integrability of the resulting functional differential equations, the expansion
(\ref{Expansion}) offers an alternative method for constructing the deformed
duality invariant action order by order as a formal power series.



\begin{thebibliography}{10}

\bibitem{CremmerJulia}
  E.~Cremmer and B.~Julia,
``The $SO(8)$ supergravity,''
  Nucl.\ Phys.\  B {\bf 159}, 141 (1979).


\bibitem{GZ}
  M.~K.~Gaillard and B.~Zumino,
``Duality rotations for interacting fields,''
  Nucl.\ Phys.\  B {\bf 193}, 221 (1981).

\bibitem{AFZ}
  P.~Aschieri, S.~Ferrara and B.~Zumino,
  ``Duality rotations in nonlinear electrodynamics and in extended
  supergravity,''
  Riv.\ Nuovo Cim.\  {\bf 31}, 625 (2008)
  \eprintN{0807.4039}.

\bibitem{Kallosh}
  R.~Kallosh,
 ``$E_{7(7)}$ symmetry and finiteness of $\N=8$ supergravity,''
  \eprintN{1103.4115}.

\bibitem{BHN}
  G.~Bossard, C.~Hillmann and H.~Nicolai,
  ``$E_{7(7)}$ symmetry in perturbatively quantised $\N=8$ supergravity,''
  JHEP {\bf 1012}, 052 (2010)
  \eprintN{1007.5472}.

\bibitem{Christian}
  C.~Hillmann,
  ``$E_{7(7)}$ invariant Lagrangian of $d=4$ $\N=8$ supergravity,''
  JHEP {\bf 1004}, 010 (2010) 
  \eprintN{0911.5225}.

\bibitem{HL} P.S.~Howe and U.~Lindstr\"{o}m,
   ``Higher order invariants in extended supergravity",
   Nucl.\ Phys.\ {\bf B181} (1981) 487


\bibitem{REK} R.E.~Kallosh,
  ``Counterterms in extended supergravities",
  Phys.\ Lett.\ {\bf B99} (1981) 122

\bibitem{HST} P.S. Howe, K.S. Stelle and P.K. Townsend,
 ``Superactions",
  Nucl.\ Phys.\ {\bf B191} (1981) 445

\bibitem{Kallosh1}
R. Kallosh, 
"$N=8$ counterterms and $E_{7(7)}$ current conservation",
JHEP{\bf 1106},073 (2011)
\eprintN{1104.5480}

\bibitem{Freedman}
  D.~Z.~Freedman and E.~Tonni,
  ``The $D^{2k} R^4$ invariants of $\N=8$ supergravity,''
  JHEP {\bf 1104}, 006 (2011)
  \eprintN{1101.1672}.
 
\bibitem{Elvang}
  H.~Elvang and M.~Kiermaier,
  ``Stringy KLT relations, global symmetries, and $E_{7(7)}$ violation,''
  JHEP {\bf 1010}, 108 (2010)
  \eprintN{1007.4813}.
  
\bibitem{BHS}
  G.~Bossard, P.~S.~Howe and K.~S.~Stelle,
  ``On duality symmetries of supergravity invariants,''
  JHEP {\bf 1101}, 020 (2011)
  \eprintN{1009.0743}.

\bibitem{Green}
 M.B.~Green, J.G.~Russo and P.~Vanhove,
  ``Automorphic properties of low energy string amplitudes in
   various dimensions'', Phys.\ Rev.\ {\bf D81}, 086008 (2010) 
   \eprintN{1001.2535}.

\bibitem{dWN} 
B.~de Wit and H.~Nicolai, 
"$N=8$ supergravity",
Nucl. Phys. {\bf B208} (1982) 323

  \bibitem{HenneauxChiral}
  M.~Henneaux and C.~Teitelboim,
  ``Dynamics of chiral (selfdual) $p$-forms,''
  Phys.\ Lett.\  B {\bf 206}, 650 (1988).
  
  
\bibitem{HenneauxCoh}
  G.~Barnich, F.~Brandt and M.~Henneaux,
  ``Local BRST cohomology in the antifield formalism. 1. General theorems,''
  Commun.\ Math.\ Phys.\  {\bf 174}, 57 (1995)
  \eprint{hep-th/9405109}.


\bibitem{WittenGaume}
  L.~Alvarez-Gaume and E.~Witten,
  ``Gravitational anomalies,''
  Nucl.\ Phys.\  B {\bf 234}, 269 (1984).

\bibitem{Singer}
  O.~Alvarez, I.~M.~Singer and B.~Zumino,
  ``Gravitational anomalies and the family's index theorem,''
  Commun.\ Math.\ Phys.\  {\bf 96}, 409 (1984).

\bibitem{Bern}
   Z. Bern, J.J. Carrasco, L.J. Dixon, H. Johansson and R. Roiban,
   ``The ultraviolet bahavior of $N=8$ supergravity at four loops",
   Phys.\ Rev.\  Lett.\  {\bf 103} (2009) 081301
   \eprintN{0905.2326}.
  
  





\end{thebibliography}
 \end{document}